%% file: main.tex
\documentclass[sigconf, reviewss]{acmart}
\acmConference[ESEC/FSE 2020]{The 28th ACM Joint European Software Engineering Conference and Symposium on the Foundations of Software Engineering}{8 - 13 November, 2020}{Sacramento, California, United States}
\acmConference[]{}{}{}

\renewcommand\footnotetextcopyrightpermission[1]{} % removes footnote with conference information in first column
\usepackage[utf8]{inputenc}

\input{headers}

\title{GraphRepo: Fast Exploration in Software Repository Mining}
\author{Alex Serban}
\email{a.serban@cs.ru.nl}
\affiliation{%
  \institution{ICIS, Radboud University}
  \institution{Software Improvement Group}
  \city{Amsterdam}
  \state{The Netherlands}
}

\author{Magiel Bruntink}
\email{m.bruntink@sig.eu}
\affiliation{
    \institution{Software Improvement Group}
    \city{Amsterdam}
    \state{The Netherlands}
}

\author{Joost Visser}
\email{j.m.w.visser@liacs.leidenuniv.nl}
\affiliation{%
	\institution{LIACS, Leiden University}
	\city{Leiden}
	\state{The Netherlands}
}

\begin{document}


% <concept_id>10011007.10011006.10011072</concept_id>
% <concept_desc>Software and its engineering~Software libraries and repositories</concept_desc>

% \ccsdesc[500]{Software and its engineering~Software libraries and repositories}

\begin{abstract}
Mining and storage of data from software repositories is typically done on a per-project basis, where each project uses a unique combination of data schema, extraction tools, and (intermediate) storage infrastructure.
We introduce GraphRepo, a tool that enables a unified approach to extract data from Git repositories, store it, and share it across repository mining projects.
GraphRepo uses Neo4j, an ACID-compliant graph database management system, and allows modular plug-in of components for repository extraction (drillers), analysis (miners), and export (mappers). 
The graph enables a natural way to query the data by removing the need for data normalisation.
GraphRepo is built in Python and offers multiple ways to interface with the rich Python ecosystem and with big data solutions.
The schema of the graph database is generic and extensible.
Using GraphRepo for software repository mining offers several advantages versus creating project-specific infrastructure: (i) high performance  for short-iteration exploration and scalability to large data sets (ii) easy distribution of extracted data  (e.g.,\ for replication) or sharing of extracted data among projects, and (iii) extensibility and interoperability.
A set of benchmarks on four open source projects demonstrate that GraphRepo allows very fast querying of repository data, once extracted and indexed.
More information can be found in the project's documentation (available at \url{https://tinyurl.com/grepodoc}) and in the project's repository (available at \url{https://tinyurl.com/grrepo}). A video demonstration is also available online (\url{https://tinyurl.com/grrepov}).
   
\end{abstract}

\maketitle

\input{acronyms}

\section{Introduction}
\label{sec:introduction}

\ac{MSR} has become an invaluable source of data for empirical software engineering research.
The tool support for \ac{MSR} evolved over time, with recent developments aiming to simplify the interactions with software repositories using clean interfaces~\cite{spadini2018pydriller}, and to evolve previous tools which used \acp{DSL}~\cite{dyer2013boa}.
These developments enable faster experimentation and facilitate integration with other tool stacks, such as statistical analysis or machine learning packages.

In this paper we introduce GraphRepo, a tool that builds on previous work by recreating the graph structure of Git repositories and by storing it in Neo4j; a database management engine which is ACID compliant.
GraphRepo brings several benefits over previous tools.
Firstly, the database engine is used for improved performance (using indexes), which facilitates fast and interactive exploration of Git repositories.
Secondly, the graph structure allows for a natural way to query repositories due to a close match to Git's internal concepts such as branching or commit-parent linkage.
Thirdly, GraphRepo offers support for working with large data sets and enables data updates with low overhead.

GraphRepo is suitable for scenarios in which Git repositories need to be explored interactively or in multiple analysis sessions, allowing for quick data updates in between sessions.
In research, GraphRepo can store large collections of Git repositories efficiently, while maintaining data freshness and performance with little overhead.
In industry, GraphRepo enables teams of developers to link and analyse their repositories, and extract (cross project) insights.

GraphRepo is built in Python and offers multiple ways to interface with the Python rich ecosystem. 
For example, it provides an \ac{ORM} interface from Neo4j to Python, such that all analyses can be written in Python, without knowledge of Neo4j's Cypher querying language.
Moreover, it provides an interface to big data solutions such as Apache Spark, allowing intense processing to be delegated from Neo4j to other tools (thus improving scalability for large-scale experiments).

This paper is organized as follows.
Initially, we discuss the tool's architecture (Section~\ref{sec:architecture}) and the graph schema (Section~\ref{sec:data_structure}).
Afterwards, we present a set of run-time benchmarks, by running GraphRepo on four medium to large projects, and illustrate a diverse set of sample queries (Section~\ref{sec:benchmarks}).
We conclude with a discussion, related and future work (Section~\ref{sec:discussion}).

\section{GraphRepo Architecture}
\label{sec:architecture}

\begin{figure}[t]
    \centering
    \includegraphics[width=7.5cm, keepaspectratio]{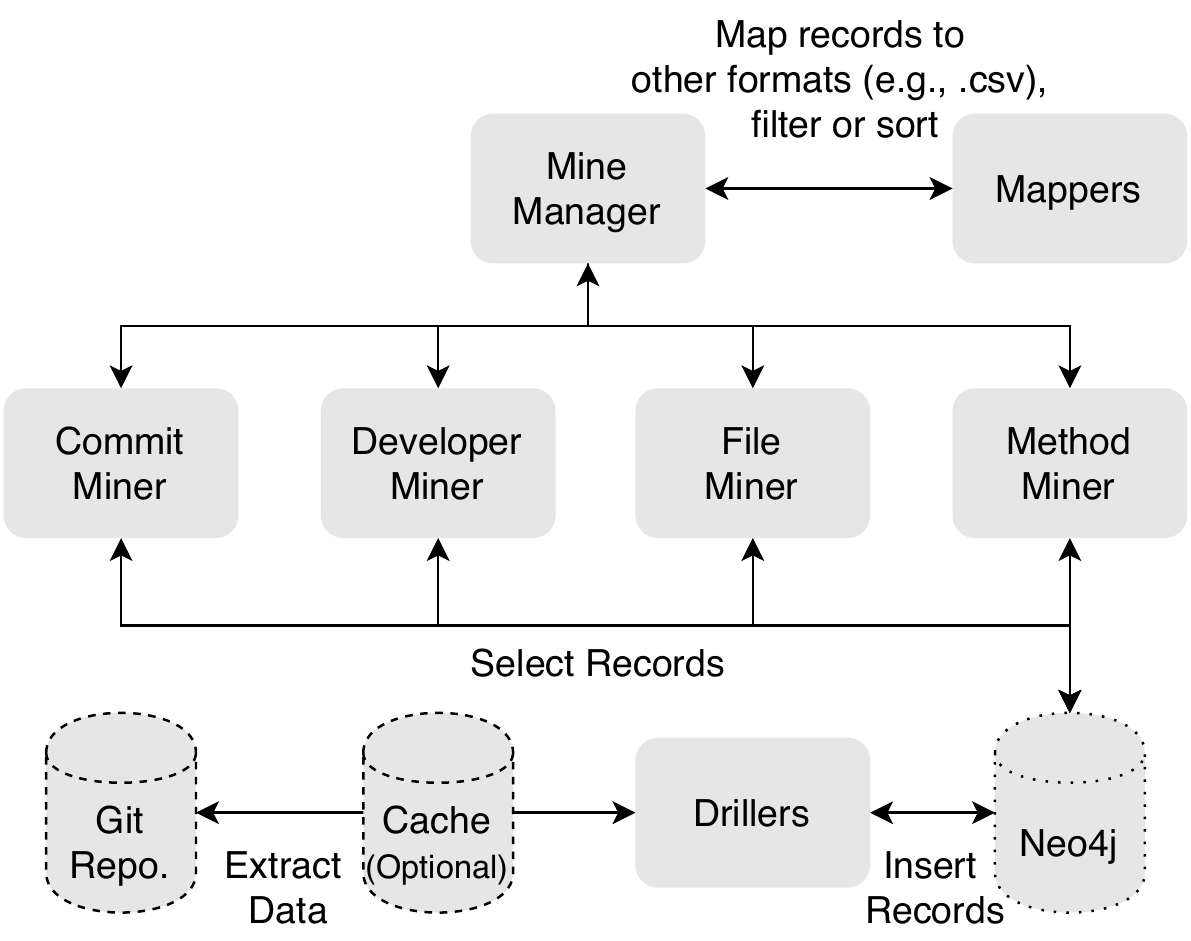}
     \caption{GraphRepo architecture.}
     \label{fig:arch}
\end{figure}

GraphRepo consists of three types of components, as illustrated in Figure~\ref{fig:arch}. 
From bottom to top, the components are (1) Drillers~--~components responsible for extracting information from Git repositories and subsequently indexing the information in Neo4j, (2) Miners~--~components responsible for querying data from Neo4j, and (3) Mappers~--~components responsible for further processing, such as mapping query results into output formats.
All components are configured using a general yaml file, as described in the project documentation\footnote{\label{grdoc}GraphRepo documentation: \url{https://tinyurl.com/grepodoc}}. In the following, we discuss the various component types in more detail.

% \paragraph{\textbf{Driller.}}
\textbf{Drillers.}
The Driller components extract data from Git repositories, including their graph structure, and insert it in Neo4j.
Under the hood, Drillers are built on PyDriller, a Python framework for \ac{MSR}~\cite{spadini2018pydriller}. PyDriller provides extraction of all core Git data, such as commits, developers, diffs, and so on. In addition, PyDriller also offers source code quality metrics through the application of the open source tool lizard~\cite{lizard}.
%We found it futile to develop another extraction framework, since PyDriller offers a broad range of features and preferred to contribute to PyDriller whenever it needed changes to accommodate our needs.

The main parameter of the Drillers consists of the batch size used to insert data in Neo4j.
The batch size allows a trade-off between performance and hardware resources, \ie~by selecting a smaller batch size the Neo4j's memory and CPU requirements are decreased at the cost of decreased performance on insert, and vice versa.
Moreover, the Drillers implement a caching mechanisms to further tweak performance and trade-off resources.

% \paragraph{\textbf{Miners.}}
\textbf{Miners.}
Miners are default components used to query the database and extract information.
They provide a set of default queries and the possibility to extend them using Python or Neo4j's Cypher query language.
At the moment there are 4 default Miners, one for each node type in the graph schema, which
provide queries on the nodes or their incoming and outgoing edges.
All Miners are coordinated by a Mine Manager, which provides a light interface for instantiation and configuration.
The use of a Mine Manager is recommended, even in case only one Miner configured, as little to no performance overhead is incurred.

% \paragraph{\textbf{Mappers}}
\textbf{Mappers.}
Mappers are custom components that can be used to further process information extracted by Miners, \eg~by sorting, filtering or converting it to other formats.
Mappers are also designed as an interface to external data processing tools (\eg~Apache Spark or Scikit-learn), in order to provide extensibility and interoperability for GraphRepo.
Therefore, Mappers facilitate scalability and enable the development of advanced data pipelines.
We also envision Mappers as a way to support experiment reproducibility, \ie~all experiments with GraphRepo can be reproduced by only sharing the configuration files and any custom Mapper.

\section{Graph Schema}
\label{sec:data_structure}

\begin{figure}[t]
    \centering
    \includegraphics[width=7.5cm, keepaspectratio]{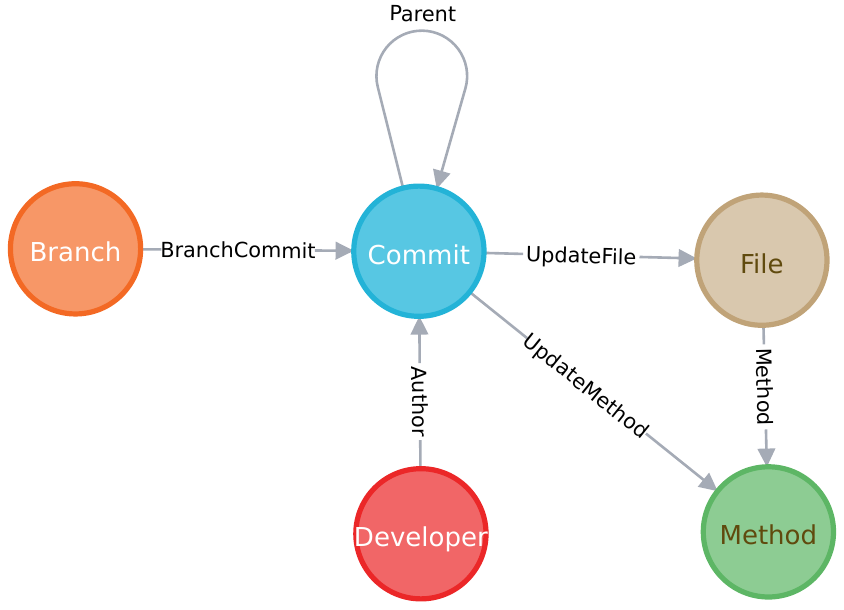}
     \caption{GraphRepo schema.}
     \label{fig:ds}
\end{figure}

The information extracted by Driller components is stored in a Neo4j graph database according to  the schema in Figure~\ref{fig:ds}.
In order to represent teams of developers that work on multiple projects, GraphRepo uses a single tenant database and adds unique project identifiers for each repository.
This allows information from different projects to be merged and queried using the same database. 
% For example, one can use the same database to measure the complexity of the code pushed by a developer in two projects with distinct technologies, in order to assess her competency per technology. 

% \paragraph{\textbf{Nodes.}}
\textbf{Node types.}
The graph node types represent the four entities universally present in software repositories: the Developers, Commits, Files and Methods.
Packages can be reconstructed from the file paths, stored as properties of the File nodes.
The decision to not represent packages as nodes was made because the package reconstruction is not universal.
Additionally, branches are represented as independent nodes, allowing faster querying and selection by branch.
Note that Git does not always allow the branch information to be reconstructed, because after merging pull requests the old branches are deleted.
As can be seen in Figure~\ref{fig:ds}, the granularity of the data is down to method level, and each method is linked to the commit that updated it.
The metadata regarding the updates is stored in the relationships properties, as discussed below.

\input{tables/projects}

\input{tables/benchmarks}

% \paragraph{\textbf{Edges (Relationships).}}
\textbf{Edge (Relationship) types.}
The edge types represent the connections (\eg~from files to methods) and the interactions (\eg~from developers to commits) in a repository, and allow to reconstruct the history of a project.
The update edge type properties hold metadata about the changes made in a commit, consisting of low level code properties, such as a file's source code before and after a commit, and high level properties such a method's complexity or fan-in measurements.
Therefore, by querying the update edges, one can extract the information about the evolution of a file or of a method. % or about the commits made by a developer.

An up-to-date list of node and edge properties is maintained in the project documentation\footnoteref{grdoc}.

\section{Benchmarks}
\label{sec:benchmarks}

We benchmark GraphRepo using four medium to large open source projects.
In all cases, we insert into Neo4j all data that can be extracted using the default Driller component, without any caching.
This data includes, for each commit, the full source code of the files changed in the commit before and after the change, together with the commit diffs.
The source code is indexed in text, as a property of the UpdateFile relationship (Figure~\ref{fig:ds}).
For most applications where the commit diffs suffice, the source code before and after each commit is redundant and adds a lot of complexity.
However, the code may be relevant for some text based applications such as source code search~\cite{niu2017learning}.
Therefore, for benchmarking purposes we index all data that can be extracted from repositories.

The experiments ran in a cloud setting, where the database was deployed using Docker on an instance with 2 virtual CPUs and 4GB RAM memory, and accessed through Neo4j's Bolt API.
Indexing was performed with a batch size of 50 inserts per batch.
The data extraction step ran on a separate instance and the two communicated over TLS.
This setting represents the realistic scenario in which a database is deployed separately from an application and can be managed and scaled independently.

\textbf{Projects.}
We used for benchmarking four open source projects which are actively maintained and developed.
The aim was to cover a broad range of popular programming languages and technologies, and explore both traditional and newly emerging projects.
The list of projects and their description is introduced in Table~\ref{tbl:project_desc}. 
The more traditional projects are Hadoop, Kibana and Tensorflow, while Jax is a recent projects which already has a mature ecosystem.

We observe that, within the dates selected, all projects have a substantial number of contributions.
Tensorflow is the most active project, with \numprint{11403} commits from 568 developers in only 5 months.
The last two columns in Table~\ref{tbl:project_desc} represent the final number of nodes and edges (relationships) for each project.
\input{tables/queries}
% Jax has an order of magnitude fewer relationships than other projects; which is not surprising given the project's age and size.

\textbf{Queries.} 
We have selected 5 queries with increasing difficulty, which need information from multiple edge hops.
In all cases, the queries can be composed from the miner components (Figure~\ref{fig:arch}), by calling one or multiple methods.
For each query, we introduce in Table~\ref{tbl:queries} the description and the number of lines of code needed to configure GraphRepo, assemble the query from one or multiple miners and map the results.

The most simple query (Q1) selects all nodes and relationships for a project and returns them as lists.
Following up (Q2) is a query that uses information from one edge type, and selects all incoming UpdateFile edges to a file, from which it extracts the \#loc.
Next, a query (Q3) that selects all the methods in a file (using the Method relationship) and, for each method, selects the incoming UpdateMethod edges from which it uses the complexity property.
The last two queries require multiple edge hops, and select the files edited by a developer (Q4) using the (Developer)~-~Author~->~(Commit)~-~UpdateFile~->~(File) path, and all methods edited by a developer (Q5) using the (Developer)~-~Author~->~(Commit)~-~UpdateMethod~->~~(Method) path.
Both queries are available in the Developer Miner component (Figure~\ref{fig:arch}) and use Neo4j's Cypher query language.

All queries require less than 10 lines of Python code to initialize the miners, assemble the queries and map the results.
Q3 requires a for-loop to get the history of each file and is slightly more complex to assemble, while Q5 requires an average over the results.

\textbf{Benchmark Results.}
For all queries and projects we present in Table~\ref{tbl:benchmarks} the time needed to process the query and return the results.
For simplicity, the time measurements are rounded half up for seconds and milliseconds.
Since the time units are small, the loss of precision after rounding is negligible. 
Additionally, we introduce the time needed to insert and index all data in Neo4j (column 3), the most costly query (column 4) and, for each query, the number of nodes or edges that have to be processed.

For queries which required a specific node (\ie~Q2, Q3, Q4, Q5) we selected the nodes with the highest number of edges in order to report the worst case performance.
For example, for Q2 we selected the files that have the most updates, for Q4 we selected the developer that edited the most files and for Q5 we selected the developer that edited the most methods.

The Driller time represents the time needed to extract the data from a Git repository and it is only informative since, for this version of GraphRepo, the Driller time depends on PyDriller.
PyDriller is currently single-thread and does not use all resources available.
A multi-thread Driller is considered for future work.

We observe that, on average, the insert time is around half the Driller time (Table~\ref{tbl:benchmarks}) and the most costly query is always for the UpdateFile relationships.
We remind that the UpdateFile edge connects a commit to a file and holds the file source code before and after a commit.
The time needed to insert these edges accounts for, on average,  66\% of the total insert time.
In case we opt out of inserting the source code and resume to the commit diff, a performance improvement of (on average) 90\% on insert time of the UpdateFile can be obtained, reducing the insert time considerably.

All queries require a modest processing time, with the most intensive query being Q3 for P4; selecting all updates of 208 methods in 5s. 
This query is particularly slower than others, \eg~slower than Q5 which also selects the method updates for approximately 75x more methods for P4, because it is implemented in two steps.
Firstly, all methods are selected and secondly, in a for loop, the updates for each method are retrieved.
Therefore, the number of queries and API calls to Neo4j is higher.
In turn, Q5 is implemented in one Cypher query and requires only one API call.
This example illustrates a trade-off between using less domain knowledge (in this case the Cypher query language) at the cost of decreased performance and vice versa.
All queries require modest processing times and allow fast and interactive exploration of software repositories.
More details and examples are available in the project's repository\footnote{GraphRepo Github repository:  \url{https://tinyurl.com/grrepo}}.

\section{Discussion and Concluding Remarks}
\label{sec:discussion}

We introduced GraphRepo, a tool for \ac{MSR} that recovers the graph structure of software repositories and stores it in Neo4j.
After inserting the data in Neo4j, GraphRepo allows interactive exploration and experimentation with software repositories.
Moreover, GraphRepo allows modular plug-in of components for data extraction, analysis and export.
The benchmarks in Section~\ref{sec:benchmarks} show that the processing time for complex queries is insignificant compared to the time needed to extract the data from repositories. % (Driller Time vs query times in Table~\ref{tbl:benchmarks}).

Nonetheless, the overhead for indexing the data in Neo4j is not admissible for all scenarios in \ac{MSR}.
GraphRepo is relevant for use-cases where the software repositories are re-used for multiple analyses which are not designed upfront, where the data is constantly updated, maintained fresh and shared in a consistent format or when interactive exploration of software repositories is desired.
For example, consider that after designing and running an analysis to answer a query (\eg~Q2, Table~\ref{tbl:queries}), one wants to answer another query for which the data is not included in the previous analysis (\eg~for Q5, Table~\ref{tbl:queries}).
In such scenarios, the data extraction step has to run again, for each new analysis. 
The performance improvement becomes more relevant when the analyses run at scale, for a large number of projects, which is the typical use case in \ac{MSR}.
The interactive exploration of software repositories enabled by GraphRepo is not enabled by other tools.
GraphRepo also creates new abstractions over classical tools for \ac{MSR}, simplifies the interface with software repositories and is built to integrate easily with other tool stacks that enable fast experimentation at scale.

\textbf{Related work.}
To extract data from Git repositories, GraphRepo uses PyDriller~\cite{spadini2018pydriller}. GraphRepo can be seen as an extension of PyDriller that provides graph persistence in Neo4j and more mining capabilities, through a lighter interface.
Using Miners and Mappers (Section~\ref{sec:architecture}), GraphRepo makes one step further to enhance extensibility, interoperability and integrate the \ac{MSR} process in advanced data pipelines.
In turn, PyDriller is built on GitPython~\cite{gitpython}, which provides better performance at the cost of increased complexity~\cite{spadini2018pydriller}.
Boa~\cite{dyer2013boa} is another tool for \ac{MSR} based on a domain specific language and with a limit on processing Java projects. 
% Compared to GraphRepo, Boa does not provide graph persistence or the flexibility enabled by Python and its rich ecosystem.

Graal~\cite{cosentino2018engineering}, and the Arthur extension~\cite{arthur} are the closest related tool stacks. 
Arthur can index the data extracted with Graal in Elasticsearch.
Compared to Neo4j, Elasticsearch is a storage engine for text search (\ie~for creating advanced inverted indexes or tokenisers) and lacks the graph query capacity of Neo4j.
Nonetheless, Arthur has advanced scheduling capacities.
Coming~\cite{arXiv-1810.08532} and Parceval~\cite{duenas2018perceval} are tools that can extract more data from repositories (\eg~commit patterns, issues) and can be used to enrich the data in GraphRepo.

GHTorrent~\cite{gousios2013ghtorent} is an offline mirror of Github which indexes social events, such as the participants in a pull request or in an issues.
Compared with GraphRepo, GHTorrent does not use data related to code changes in commits.
Nonetheless, the social interactions from GHTorrent can be used to complement the data used by GraphRepo and their integration is an interesting avenue for future work. 

% \todo{Awesome Empirical Software Engineering list~\cite{awesome-msr}}.

% \todo{Graal~\cite{cosentino2018engineering}}.

% \todo{Coming~\cite{arXiv-1810.08532}}.

\textbf{Future work.}
The main performance challenge of GraphRepo comes from extracting data from Git repositories.
The Driller components are currently single-thread and a multi-thread driller is considered for future work.
This component may divide a software repository in several chunks (\eg~by date) and process them in parallel.
The results could be cached using the already available caching mechanisms, ordered and stored in Neo4j.
Other performance improvements may come from adding better Neo4j indexes by default, which in turn can improve the insert and query times.

More miners and mappers, which can enhance the default querying, extensibility and interoperability capacity of GraphRepo are also planned.
We also welcome contributions through Github, and look forward to requests and use-cases from the community.

\bibliographystyle{ACM-Reference-Format}
\bibliography{bibliography}

% \appendix

% \section{Demonstration Resources}

% \subsection{Walk-through}

% The demo consists of indexing a repository using GraphRepo and exploring the data interactively, using a multitude of queries.
% The complete demo (walk-through) is public on Github (\url{https://tinyurl.com/grwalkthrough}), and consists of a Jupyter Notebook.
% The queries are default, from the Miner components (Figure~\ref{fig:arch}) and show the light interfaces GraphRepo exposes to Git repositories.
% A default Mapper (Figure~\ref{fig:arch}) is also demonstrated.

% In sequential order, the demo consists of:
% \begin{itemize}
%     \item Discuss the purpose of GraphRepo.
%     \item Showcase the architecture.
%     \item Index a repository in Neo4j using GraphRepo (using the Driller components).
%     \item Perform multiple queries using the Miners.
%     \item Demonstrate a default Mapper.
%     \item Reference public materials.
% \end{itemize}

% \subsection{Tool and Data Availability}
% The tool is open source and has been already used in research and industry, in the context of Software Improvement Group, Radboud University and Leiden University.

% The online materials can be accessed using the following links:
% \begin{itemize}
%     \item Online documentation: \url{https://tinyurl.com/grepodoc}
%     \item Project's Github: \url{https://tinyurl.com/grrepo}
%     \item Screencast: \url{https://tinyurl.com/grrepov}
%     \item Demo materials: \url{https://tinyurl.com/grwalkthrough}
% \end{itemize}

\end{document}

%% file: headers.tex
\usepackage{acronym}
\usepackage{xcolor}

\usepackage{caption}
\usepackage{subfig}
\usepackage{url}

\usepackage{numprint}
\input{custom_commands}

\usepackage[export]{adjustbox}

%% file: custom_commands.tex
\newcommand{\specialcelll}[2][c]{%
  \begin{tabular}[#1]{@{}l@{}}#2\end{tabular}}

\makeatletter
\newcommand\footnoteref[1]{\protected@xdef\@thefnmark{\ref{#1}}\@footnotemark}
\makeatother

\newcommand{\ie}{i.e.,}
\newcommand{\eg}{e.g.,}

%% file: acronyms.tex
\acrodef{MSR}[MSR]{Mining software repositories}
\acrodef{ORM}[ORM]{object-relational mapping}
\acrodef{DSL}[DSL]{domain specific language}

%% file: tables/projects.tex
\begin{table*}[t]
    \caption{Projects descriptions, where the number of nodes and edges for each project is extracted from Neo4j.}
    \label{tbl:project_desc}
    \centering
    \begin{tabular}{lllcrrrrrr} 
    % \begin{tabular}{lllccccccc} 
        \toprule
         ID & Name & Technologies & Start - End Dates & \#Devs. & \#Commits & \#Files & \#Methods & \#Nodes & \#Edges  \\
        \midrule
        P1 & Hadoop & Java & 01.01.2018 - 01.01.2019 & 107 & \numprint{2359} & \numprint{6613} & \numprint{43817} &  \numprint{52897} & \numprint{127837} \\ 
        P2 & Jax & Python &  01.01.2019 - 01.05.2020 &  182  &  \numprint{3721} &  \numprint{299} &  \numprint{7963} & \numprint{12166} & \numprint{42640}  \\  
        P3 & Kibana & Java/Javascript & 01.06.2018 - 01.06.2019 & 232 & \numprint{5826} & \numprint{26827} & \numprint{15227} & \numprint{48113} & \numprint{130642}  \\        
        P4 & Tensorflow & C++/Python/Go & 01.12.2019 - 01.05.2020 & 568 & \numprint{11403} & \numprint{11549} & \numprint{51335} & \numprint{74856} &  \numprint{213368} \\ 
        \bottomrule
    \end{tabular}
\end{table*}

%% file: tables/benchmarks.tex
% \begin{table*}[th]
%     \caption{Benchmark results. The Driller column is present for informative purposes and represents the time needed to extract the data using PyDriller.}
%     \label{tbl:benchmarks}
%     \centering
%     \begin{tabular}{ccccccccc} % 
%         \toprule
%          PID & Driller & Index & Longest Index & Q1 & Q2 & Q3 & Q4 & Q5 \\
  
%         \midrule
%         P1 & 13:02.08 & 06:51.65 & UpdateFiles 4.54.90  & 00:00.21  & 62c  00:02.69   & 143m  00:03.011 & 443f  00:00.34 & 1262m  00:00.4025 \\ % hadoop
        
%         P2 & 08:53.30  & 04:57.22 & UpdateFiles - 03:39.50 & 00:00.22 & 129c 00:01.18 & 192m 00:04.16 & 60f 00:00.24 & 929m 00:00.39 \\ % jax
        
%         P3 & 35.30.33 & 13:38.31  & UpdateFiles 08:56.83  & 00:00.21  & 36c 00:00.42 & 59m 00:01.78  &  3083f 00:00.53  & 3048m 00:00.67 \\ % Kibana
        
%         P4 & 1:59:41.55 & 51:06.15 & UpdateFiles 33:31.21 &  00:01.33  &  77c 00:01.76  & 208m 00:05.80  &  4782f 00:00.95  &  15545m, 00:02.40 \\  % Tensorflow
        
%         \bottomrule
%     \end{tabular}
% \end{table*}

\begin{table*}[th]
    \caption{Benchmark results. The Driller column is present for informative purposes and represents the time needed to extract the data using PyDriller.}
    \label{tbl:benchmarks}
    \centering
    % \begin{tabular}{ccccccccccccccc} % 
    \begin{tabular}{lrcp{2em}rrrrrrrrrrr} %     
        \toprule
             &   Driller  &  Insert \& Index  & \multicolumn{2}{c}{Most Costly Insert} & \multicolumn{1}{c}{Q1} & \multicolumn{2}{c}{Q2} & \multicolumn{2}{c}{Q3} & \multicolumn{2}{c}{Q4} & \multicolumn{2}{c}{Q5}  \\
         PID & Time & Time & Query & Time & Time & \#Updates & Time & \#Methods & Time &  \#Files & Time & \#Updates & Time  \\
        \midrule
        P1 & 13m & 7m & UpdateFile & 5m  & 14s  & 62 & 3s & 143 & 3s & 443 &  34ms & \numprint{1262} & 40ms \\ % hadoop
                
        P2 & 8m  & 5m & UpdateFile & 3m & 8s & 129 & 1s & 192 & 4s & 60 & 24ms & \numprint{929} & 39ms \\ % jax
        
        P3 & 35m & 13m  & UpdateFile & 9m  & 13s  & 36 & 1s & 59 & 1s  &  \numprint{3083} & 53ms  & \numprint{3048} & 67ms \\ % Kibana       

        P4 & 1h59m & 51m & UpdateFile & 34m & 21s & 77 & 2s  & 208 & 5s & \numprint{4782} & 95ms  &  \numprint{15545} & 2s \\  % Tensorflow        
        \bottomrule
    \end{tabular}
\end{table*}

% \begin{table*}[th]
%     \caption{Benchmark results. The Driller column is present for informative purposes and represents the time needed to extract the data using PyDriller.}
%     \label{tbl:benchmarks}
%     \centering
%     % \begin{tabular}{ccccccccccccccc} % 
%     \begin{tabular}{lrcp{2em}rrrrrrrrrrr} %     
%         \toprule
%              &   Driller  &  Insert \& Index  & \multicolumn{2}{c}{Most Costly Insert} & \multicolumn{1}{c}{Q1} & \multicolumn{2}{c}{Q2} & \multicolumn{2}{c}{Q3} & \multicolumn{2}{c}{Q4} & \multicolumn{2}{c}{Q5}  \\
%          PID & Time & Time & Query & Time & Time & \#Updates & Time & \#Methods & Time &  \#Files & Time & \#Updates & Time  \\
%         \midrule
%         P1 & 13m & 2m & UpdateFile & 1m  & 14s  & 62 & 3s & 143 & 3s & 443 &  34ms & \numprint{1262} & 40ms \\ % hadoop
                
%         P2 & 8m  & 1m & UpdateFile & 30s & 8s & 129 & 1s & 192 & 4s & 60 & 24ms & \numprint{929} & 39ms \\ % jax
        
%         P3 & 35m & 4m  & UpdateFile & 3m  & 13s  & 36 & 1s & 59 & 1s  &  \numprint{3083} & 53ms  & \numprint{3048} & 67ms \\ % Kibana       

%         P4 & 1h59m & 7m & UpdateFile & 5m & 21s & 77 & 2s  & 208 & 5s & \numprint{4782} & 95ms  &  \numprint{15545} & 2s \\  % Tensorflow        
%         \bottomrule
%     \end{tabular}
% \end{table*}

%% file: tables/queries.tex
\begin{table}[th]
    \caption{Query description and complexity in nr. of lines of code (\#loc) for benchmark queries. The \#loc includes both the initialization of GraphRepo and the result mapping.}
    \label{tbl:queries}
    \centering
    \begin{tabular}{clc} % 12
        \toprule
         ID & Description & \#loc \\
        \midrule
        Q1 & \specialcelll{Select all nodes and relationships for a project}. & 2 \\ 
        Q2 & \specialcelll{Select the evolution of a file's \#loc over time, \\ for a specific file.} & 3 \\ 
        Q3 & \specialcelll{Select the evolution of method complexity \\ over time, for all methods in a file.} & 7 \\ 
        Q4 & \specialcelll{Select all files edited grouped by file type, \\ for a specific developer.} & 4 \\  
        Q5 & \specialcelll{Select the average complexity of methods edited \\ by a developer in all her commits.} & 4\\
        \bottomrule
    \end{tabular}
\end{table}